\documentstyle[11pt,newpasp,twoside]{article}
\def\lesssim{{_ <\atop{^\sim}}}

\def\edcomment#1{\iffalse\marginpar{\raggedright\sl#1\/}\else\relax\fi}
\reversemarginpar

\begin{document}
\title{Substructure and halo density profiles in a Warm Dark 
Matter Cosmology}
\author{Pedro Col\'{\i}n\altaffilmark{1}, Vladimir Avila-Reese\altaffilmark{1}, 
and Octavio Valenzuela\altaffilmark{2}}
\altaffiltext{1}{Instituto de Astronom\'\i a-UNAM, A.P. 70-264, 04510 
M\'exico, D. F.}
\altaffiltext{2}{Astronomy Department, NMSU, Box 30001, 4500, Las Cruces,
NM 88003-0001}

\begin{abstract}
High-resolution cosmological N-body simulations were performed in 
order to study the substructure of Milky Way-like galactic halos and
the density profiles of halos in a WDM scenario. The results
favor this scenario with respect to the CDM one.
\end{abstract}

$\bullet${\bf The warm dark matter (WDM) scenario} has been proposed 
as an alternative to alleviate
the overly number of small halos (satellites) found within  Milky 
Way-size halos in cold dark matter (CDM) N-body simulations with 
respect to observations (Hogan \& Dalcanton 2000; White \& Croft 2000). 
We have performed cosmological N-body simulations with several 
free-streaming scale-lengths, R$_f=0.2$, 0.1 and 0.05 Mpc, 
corresponding to "warmon" masses m$_W$ of 1.7, 1.0 and 0.6 KeV.
The power spectrum of fluctuations below R$_f$ is exponentially
damped. We have used a flat cosmological model with $\Omega_ {\Lambda}=
h=0.7$; this is one of the most successful models at large scales. 
Relatively isolated (host) halos of masses of the 
order of the Milky-Way or Andromeda galaxies were chosen to be
re-simulated within a multiple mass scheme. In this way, our host 
halos had more than $10^5$ particles each and the abundance 
of sub-halos within them could be studied with confidence. 

$\bullet${\bf Abundance of satellite halos.}
Despite power spectrum is suppressed below the free-streaming scale 
R$_f$, halos close or smaller than R$_f$ do form.  
Comparing the observed cumulative maximum circular 
velocity satellite function with models and in the understanding 
that a galaxy will form in each halo, we find that models with 
R$_f\sim 0.1-0.2$ Mpc are preferred (Col\'{\i}n et al. 2000); 
in these cases the number of Milky Way satellites is $\sim 2-3$ 
times less than in the CDM simulation (see also Klypin et al. 1999).
We find that in the WDM scenario not only less small halos are 
formed than in the CDM one but also are the small halos more easily 
disrupted during host halo evolution. 
The percentage of halo destruction for the CDM and WDM (R$_f=0.1$ Mpc) 
simulations, from $z=1$ to $z=0$ is 35\% and 63\%, respectively. 
Interestingly, in the CDM case, the rates 
of satellite accretion and destruction reach a balance in such a 
way that the number of guest halos remains almost constant
since $z=1$ (see also Moore et al. 1999). The more efficient disruption of
halos in the WDM scenario is likely due to the fact that small
halos are less concentrated. 

$\bullet${\bf Density profiles.} The host WDM halos (M$\gg$M$_f$, where
M$_f$ is the mass corresponding to R$_f$) have 
density profiles and concentrations similar to those of the CDM 
halos. The guest WDM halos (M$\lesssim$ M$_f$) are systematically
less concentrated than the corresponding CDM halos. Note that the
cosmogony of these halos is not hierarchical but they form
through fragmentation and monolithic collapse. In the 
$10^9-10^{11}$ h$^{-1}M_{\odot}$ mass range the c$_{1/5}$ 
concentrations for the WDM (R$_f=0.2$ Mpc) model are $\approx 1.8-1.2$
times smaller than those obtained in CDM simulations. The 
c$_{1/5}$ parameter is defined as the ratio between the halo 
radius and the radius where 1/5 of the halo mass is contained. 
In order to explore with more detail the inner structure
of halos with masses below M$_f$, we have increased R$_f$ and 
studied host halos which have several times $10^4$ particles. The results
confirm that the concentration of halos with M$\lesssim$M$_f$ is 
smaller than the one corresponding to CDM halos, and this
difference can hardly be explained as a result of statistical 
scatter. On the other hand, the density profiles of most 
halos do not strongly depart from the NFW profile; the inner 
slopes are $\sim -1$, shallower than for the CDM halos. Finally, 
in our new simulations, we have introduced the corresponding thermal 
velocity dispersions $v_{\rm th}$ and have found that the structure 
of the halos is not affected, even with a $v_{\rm th}$ 2 times larger
than the corresponding $v_{\rm th}$ for m$_W\sim 0.6-1$ KeV (making 
echo to claims that a larger $v_{\rm th}$
could be possible if warmons self-interact weakly, Hannestad 2000). 

$\bullet${\bf In conclusion}, the WDM scenario offers a viable 
solution to the satellite overabundance problem and it improves
the agreement with observations regarding the concentrations and
the inner density profiles of dwarf and LSB galaxies (according to 
recent observational studies, the halo structure of these galaxies
does not disagree with the NFW profile, although their concentrations
are smaller than in the CDM case, e.g. van den Bosch \& Swaters 2000).
Besides, the formation of realistic disks seem to be favored
in the WDM scenario (Sommer-Larsen \& Dolgov 2000).
On the other hand, for warmon masses $>0.75$ KeV, the power 
spectrum of the Ly$\alpha$ clouds 
agrees with observations (Narayanan et al. 2000).
For these WDM models, we do not find serious difficulties regarding
the Tully-Fisher relation and the dwarf satellite formation epochs
(the power spectrum regeneration is very quick and efficient,
e.g., White \& Croft 2000). Nevertheless, the WDM scenario
could not explain shallow halo cores and central densities
independent of mass; if observations confirm these trends
(e.g., Firmani et al. 2000) the WDM scenario should be 
abandoned.  
 
V.A. received financial support from CONACyT grant 27752-E.

\end{document}